\documentclass[twocolumn,superscriptaddress,letter]{revtex4}%
\usepackage{amssymb}
\usepackage{color}
\usepackage{graphicx}
\usepackage{dcolumn}
\usepackage{bm}
\usepackage[header,title,page,titletoc]{appendix}
\usepackage{amsmath}
\usepackage{amsfonts}%
\providecommand{\U}[1]{\protect\rule{.1in}{.1in}}

\begin{document}
\title{Preparation of NOON State Induced by Macroscopic Quantum Tunneling in an Ising Chain}
\author{Chun-Li Zang}
\affiliation{Department of Physics, Beijing Normal University, Beijing 100875, China}
\author{Jing Yu}
\affiliation{Department of Physics, Liaoning Shihua University, Fushun, 113001, P. R. China}
\author{Wan-Li Yang}
\affiliation{State Key Laboratory of Magnetic Resonance and Atomic and Molecular Physics,
Wuhan Institute of Physics and Mathematics, Chinese Academy of Sciences, Wuhan
430071, China}
\author{Mang Feng}
\affiliation{State Key Laboratory of Magnetic Resonance and Atomic and Molecular Physics,
Wuhan Institute of Physics and Mathematics, Chinese Academy of Sciences, Wuhan
430071, China}
\author{Su-Peng Kou}
\thanks{spkou@bnu.edu.cn}
\affiliation{Department of Physics, Beijing Normal University, Beijing 100875, China}

\begin{abstract}
In this brief report, we propose a possible way, theoretically and
experimentally, to generate a NOON state of the two degenerate
ferromagnetic ground states of the Transverse Ising Model. In our scheme
we employ the macroscopic quantum tunneling (MQT) effect between the two degenerate
ferromagnetic ground states to realize the NOON state. Our calculation about the
MQT process is based on a higher-order degenerate perturbation method. After doing a
transformation, the MQT process could also be treated as the hopping of individual
virtual fermions in the spin chain, which will leads to an analytical description of
tunneling process. The experimental feasibility for generating the NOON state is
discussed in the setup of linear ion trap.

\end{abstract}
\author{}
\maketitle

The NOON states are a kind of special quantum states with two orthogonal
component states in maximal superposition, where one is all spin-up and the
other is all spin-down, such as $\left\vert NOON\right\rangle =\left(
\left\vert \uparrow\uparrow\cdots\uparrow\uparrow\right\rangle \pm\left\vert
\downarrow\downarrow\cdots\downarrow\downarrow\right\rangle \right)  /\sqrt
{2}$. The recent interests in production of NOON states mainly arise from the
crucial role they play in quantum optical lithography \cite{lith} and quantum
metrology \cite{metr1,metr2,metr3} as well as quantum information processing
\cite{qip}, including quantum gating \cite{gate} and precision measurement of
phases \cite{mea,mea1,mea2}. Many efforts have been devoted to preparation of
NOON states theoretically and experimentally by using trapped ions
\cite{ion1,ion2,ion3}, cold atoms \cite{cold1,cold2}, atomic ensembles
\cite{en}, superconducting quantum circuits \cite{sc}, cavity quantum
electrodynamics \cite{cav}, and linear or nonlinear optical elements
\cite{metr3,mea,lin1,lin2,lin3}.

In this Brief Report, we propose an idea to create the NOON states formed by
two degenerate ferromagnetic ground states of a transverse Ising model by the
mechanism of macroscopic quantum tunneling (MQT). In our case, the transverse
Ising model is given by
\begin{equation}
\hat{H}=\hat{H}_{0}+\hat{H}_{I}, \label{1}%
\end{equation}
where $\hat{H}_{0}=-J\sum_{i}\sigma_{i}^{z}\sigma_{i+\hat{e}_{x}}^{z}$ and
$\hat{H}_{I}=h^{x}\sum\limits_{i}\sigma_{i}^{x}$. In general, the transverse
Ising model includes two quantum limits at absolute zero temperature: the
ferromagnetic (FM) limit ($J>h^{x}$) and the paramagnetic limit ($J<h^{x}$),
separated by the strength of the transverse field $J_{c}$ \cite{sachdev}. In
the present model, $\hat{H}_{I}$ is the perturbation term to induce quantum
tunneling, as discussed later. The global phase diagram of the transverse
Ising model at zero temperature, as shown in Fig. 1, has been well known
\cite{sachdev,ising}, where quantum phase transition occurs from the
ferromagnetic state with $J>h^{x}$ to the gapped spin-polarized state with
$J<h^{x}$ at the quantum transition point $J/h^{x}=1$. As a result, for large
$J$ $\left(  J>h^{x}\right)  $, the ground state owns a spontaneous non-zero
magnetization in $z$ direction due to its FM order. For small $J$ $\left(
J<h^{x}\right)  $, the ground state is the uncorrelated spin state, also
called the spin-polarized state.

The key idea in our proposal is the quantum tunneling between degenerate
ground states induced by a transverse perturbation field. In a macroscopic
system, such a quantum tunneling is usually referred to MQT, which has
appeared in a wide range of research fields, such as quantum oscillations
between two degenerate wells of \textrm{NH}$_{3}$, quantum coherence in
one-dimensional charge density waves, the ferromagnetic single domain magnets
and biased Josephson junctions \cite{Razag}. In general, the MQT can be found
in systems with two or more separated 'classical' states that are
macroscopically distinct. For instance, the MQT in macroscopic systems at low
temperature, such as the superconducting quantum interference device, might be
feasible for observation \cite{Leggett}.

There have been some useful theoretical methods to deal with the MQT, such as
semiclassical WKB (Wentzel, Kramers and Brillouin) method \cite{landau} and
Instanton method in path integration \cite{coleman}. In the present work, we
study the MQT in a transverse Ising chain by the perturbation method
\cite{kou1,kou1',yu}, based on the fact that the transverse Ising model in a
finite size owns the degeneracy on the ground state, which can be removed by
the MQT process. This mechanism can result in production of NOON states in the
transverse Ising model.

\begin{figure}[ptb]
\includegraphics[width=0.48\textwidth]{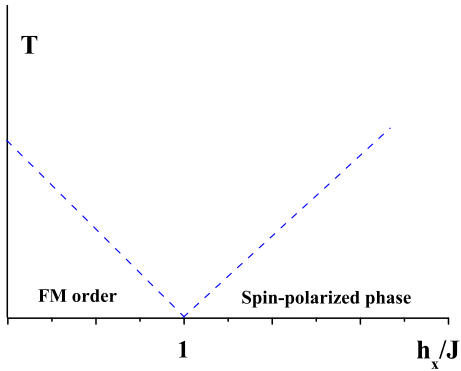}\caption{The global phase
diagram of transverse Ising model at zero temperature, where the quantum phase
transition happens as $J$ crosses the critical value $J_{c}$. Our model stays
in the ferromagnetic limit ($J\gg h_{x}$) of the phase diagram. }%
\label{Fig.1}%
\end{figure}

To classify the degeneracy of the two ground states of $\hat{H}$, we introduce
two string operators $W_{X}=\prod\limits_{i}\sigma_{i}^{x}$ and $W_{Z}
=\prod\limits_{i}\sigma_{i}^{z}$ \cite{do,kou1}, which commute with $H_{0}$,
i.e., $\left[  \hat{H}_{0},W_{X}\right]  =\left[  \hat{H}_{0},W_{Z}\right]
=0$ and are anti-commutated with each other, i.e., $\left\{  W_{Z},
W_{X}\right\}  =0$. Due to these properties, we may represent $W_{X}$ and
$W_{Z}$ by two-component pseudo-spin operators $\tau^{x}$ and $\tau^{z}$,
respectively, as $W_{X}\rightarrow\tau^{x}$ and $W_{Z}\rightarrow\tau^{z}$,
for which we have $\tau^{z}\left\vert \downarrow\downarrow\cdots
\downarrow\downarrow\right\rangle =-\left\vert \downarrow\downarrow
\cdots\downarrow\downarrow\right\rangle ,\tau^{z}\left\vert \uparrow
\uparrow\cdots\uparrow\uparrow\right\rangle =$ $\left\vert \uparrow
\uparrow\cdots\uparrow\uparrow\right\rangle $, where $\left\vert
\uparrow\uparrow\cdots\uparrow\uparrow\right\rangle $ and $\left\vert
\downarrow\downarrow\cdots\downarrow\downarrow\right\rangle $ are two
degenerate ferromagnetic many-body ground states, as show in Fig. \ref{Fig.1}.

In order to illustrate the MQT process in the Ising chain, we transform the
Ising spin chain into the system of free fermions by a perturbative method
\cite{kou1'}. In the absence of $\hat{H}_{I}$, the fermions in this model have
flat bands with the energy spectrum of $E(k)=2J$, which implies that there is
a mass gap $2J$ of the fermions in the transformed system, and the fermion
cannot move in such a case. To change this situation, we may introduce the
transverse perturbation $\hat{H}_{I}$, which drives the fermions to hop in the model.

In the perturbative approach, the perturbative operators $\sigma_{i}^{x}$ can
be represented by hopping terms of the fermions, $\sigma_{i} ^{x}%
\rightarrow(\psi_{i}^{\dag}\psi_{i+1}+ \psi_{i}^{\dag}\psi_{i+1}^{\dag}%
+h.c.)$, where $\psi_{i}^{\dag}$ ($\psi_{i}$) is the creation (annihilation)
operator of a spinless fermion at the site $i$. In addition, we introduce a
constraint for single occupation per site (i.e., hard-core constraint),
\begin{equation}
(\psi_{i}^{\dag})^{2}|\Psi\rangle=|\Psi\rangle\text{ or }\psi_{i}^{\dag}%
|\Psi\rangle=\psi_{i}|\Psi\rangle,\nonumber
\end{equation}
where $|\Psi\rangle$ denotes a many-body quantum state.

Therefore, by the perturbative method, the Hamiltonian in Eq. (1) can be
rewritten by spinless fermions as,
\begin{equation}
\hat{H}=2J\sum_{i}\psi_{i}^{\dag}\psi_{i}-h^{x}\sum_{\left\langle
i,j\right\rangle }(\psi_{i}^{\dag}\psi_{j}+\psi_{i}^{\dag}\psi_{j}^{\dagger
}+h.c). \label{2}%
\end{equation}
Under the Fourier transformation, we may assume periodic boundary condition
for the Ising chain, and obtain the dispersion of the fermion system in Eq.
(\ref{2}) as
\begin{equation}
E(k)=\sqrt{(2h^{x}\cos k+2J)^{2}+4(h^{x}\sin k)^{2}},
\end{equation}
where $k$ is the wave vector, and for simplicity we set the separation of two
nearest neighbor spins to be unity. So the energy gap of the fermions is
obtained as $m_{F}=2J\sqrt{1-\frac{h^{x}}{J}}$, and thereby we may manipulate
the dispersion of the fermions by tuning the external field $h^{x}$. In the
language of the free fermion system, during the tunneling process, a pair of
virtual fermions are created, one of which moves around the chain until
annihilation by meeting another fermion \cite{kou1,kou1',yu}.
\begin{figure}[ptb]
\includegraphics[clip,width=0.45\textwidth]{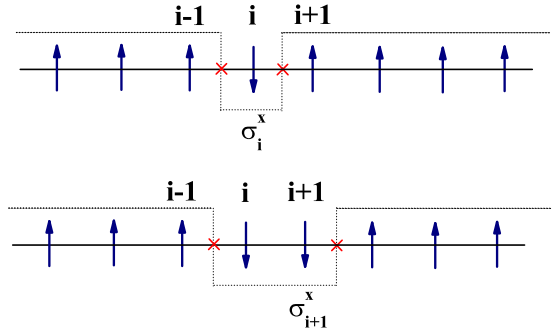}\caption{(Color online)
The hopping process of the fermions from the site $i$ to $i+1$. At site $i$, a
pair of fermions are generated by the operation $\sigma_{i}^{x}$. One fermion
(at right cross in the lower string) hops when one does an operation
$\sigma_{i+1}^{x}.$ Then the hopping process of fermions is driven by the
transverse perturbation field $h_{x}$. We represent the fermions by red
crosses on the string.}%
\label{Fig.2}%
\end{figure}

In the thermodynamic limit, the two ferromagnetic ground states of Eq.
(\ref{1}) are degenerate. However, there is very tiny energy splitting $\Delta
E$ between the two ground states in a finite size chain. Since $\Delta E$ is
much smaller than the energy gap $m_{F}$ in the fermionic excitation, we may
safely ignore those excited states and only concentrate on the MQT between the
two degenerate ground states in our perturbative calculation.

Now we employ a high-order degenerate perturbative approach to calculate the
MQT effect in our system \cite{kou1,kou1',yu}. Firstly we introduce the
transformation operator $\hat{U_{I}}(0,-\infty)$ defined in units of
$\hslash=1$,
\begin{equation}
\hat{U_{I}}(0,-\infty)=\mathrm{\hat{T}}\exp(-i\int_{-\infty}^{0}\hat{H}
_{I}^{\prime}(t^{\prime})dt^{\prime}), \label{zero}%
\end{equation}
where $\hat{H}_{I}^{\prime}(t)=e^{i\hat{H}_{0}t}\hat{H}_{I}e^{-i\hat{H}_{0}t}$
is for Heisenberg picture, and $\mathrm{\hat{T}}$ is the time-ordering
operator. The transformation operator $\hat{U_{I}}(0,-\infty)$ in Eq.
(\ref{zero}) can be further written as
\begin{equation}
\hat{U_{I}}(0,-\infty)\left\vert m\right\rangle =\sum_{j=0}^{\infty}\hat
{U}_{I}^{(j)}(0,-\infty)\left\vert m\right\rangle , \label{U0}%
\end{equation}
where
\begin{align}
\hat{U}_{I}^{(0)}(0,-\infty)\left\vert m\right\rangle  &  =\left\vert
m\right\rangle ,\nonumber\\
\hat{U}_{I}^{(j\neq0)}(0,-\infty)\left\vert m\right\rangle  &  =(\frac
{1}{E_{0}-\hat{H}_{0}}\hat{H_{I}})^{j}\left\vert m\right\rangle
\end{align}
and $\left\vert m\right\rangle =\left(
\begin{array}
[c]{c}%
\left\vert \uparrow\uparrow\cdots\uparrow\uparrow\right\rangle \\
\left\vert \downarrow\downarrow\cdots\downarrow\downarrow\right\rangle
\end{array}
\right)  $. The element of the transformation matrix from the state
$\left\vert m\right\rangle $ to $\left\vert n\right\rangle $ becomes
$\left\langle n\right\vert \hat{U_{I}}(0,-\infty)\left\vert m\right\rangle $,
corresponding to the energy,
\begin{equation}
E=\left\langle n\right\vert \hat{H}\hat{U_{I}}(0,-\infty)\left\vert
m\right\rangle =E_{0}+\delta E, \label{deltE}%
\end{equation}
with $E_{0}$ the eigenvalue of the Hamiltonian $\hat{H}_{0}$ for the
eigenstate $\left\vert m\right\rangle $.

When the MQT occurs between the two degenerate many-body ground states
$\left\vert \uparrow\uparrow\cdots\uparrow\uparrow\right\rangle $ and
$\left\vert \downarrow\downarrow\cdots\downarrow\downarrow\right\rangle $, we
may describe the process in spinless fermion language as that, a pair of
spinless fermions are created, and then one of them travels around the chain
and leads to a string operator. As a result, the dominant term in Eq.
(\ref{U0}) is labeled by $j=N-1$, where $N$ is the site number of the Ising
chain. Considering the tunneling process, we obtain the perturbative energy
as
\begin{align}
\delta E  &  =\left\langle n\right\vert \hat{H_{I}}\hat{U_{I}}(0,-\infty
)\left\vert m\right\rangle \label{sp}\\
&  =\left\langle n\right\vert \hat{H_{I}}\sum_{j=0}^{\infty}\hat{U}_{I}%
^{(j)}(0,-\infty)\left\vert m\right\rangle \nonumber\\
&  =\left\langle n\right\vert \hat{H_{I}}\hat{U}_{I}^{(N-1)}(0,-\infty
)\left\vert m\right\rangle ,\nonumber
\end{align}
where the operator $\hat{H_{I}}\hat{U}_{I}^{(N-1)}(0,-\infty)$ is proportional
to a string operator $W_{X,Z}$.

To calculate the MQT between the two ground states, we consider a virtual
fermion propagates around the chain, leading to the quantum state $\left(
\begin{array}
[c]{c}%
\left\vert \uparrow\uparrow\cdots\uparrow\uparrow\right\rangle \\
\left\vert \downarrow\downarrow\cdots\downarrow\downarrow\right\rangle
\end{array}
\right)  $ flipped as
\begin{equation}
\left(
\begin{array}
[c]{c}%
\left\vert \downarrow\downarrow\cdots\downarrow\downarrow\right\rangle \\
\left\vert \uparrow\uparrow\cdots\uparrow\uparrow\right\rangle
\end{array}
\right)  =\tau^{x}\left(
\begin{array}
[c]{c}%
\left\vert \uparrow\uparrow\cdots\uparrow\uparrow\right\rangle \\
\left\vert \downarrow\downarrow\cdots\downarrow\downarrow\right\rangle
\end{array}
\right)  .
\end{equation}
Using Eq. (\ref{sp}), we may obtain the energy shift $\delta E$ as
\begin{align}
\delta E  &  =U_{I}^{(N)}\nonumber\\
&  =\langle\uparrow\uparrow\cdots\uparrow\uparrow\mid\hat{H_{I}}(\frac
{1}{E_{0}-\hat{H}_{0}}\hat{H_{I}})^{N-1}\left\vert \downarrow\downarrow
\cdots\downarrow\downarrow\right\rangle .
\end{align}

Due to the translation invariance of Eq. (\ref{1}), we choose the site $i$ as
the starting point of the MQT process and we have
\begin{align}
&  (\frac{1}{E_{0}-\hat{H}_{0}}\hat{H_{I}})\left\vert \downarrow
\downarrow\cdots\downarrow\downarrow\right\rangle \nonumber\\
&  \rightarrow N(\frac{h^{x}}{E_{0}-\hat{H}_{0}}\sigma_{i}^{x})\left\vert
\downarrow\downarrow\cdots\downarrow\downarrow\right\rangle \nonumber\\
&  =N(\frac{h^{x}}{E_{0}-\hat{H}_{0}})\left\vert \Psi_{i}\right\rangle ,
\end{align}
where $\left\vert \Psi_{i}\right\rangle $ is the excited state of the two
fermions at the sites $i$ and $i-1$ with the energy $E_{0}+4J$. Using $\hat
{H}_{0}\left\vert \Psi_{i}\right\rangle =(E_{0}+4J)\left\vert \Psi
_{i}\right\rangle $, we have
\[
(\frac{1}{E_{0}-\hat{H}_{0}}\hat{H_{I}})\left\vert \downarrow\downarrow
\cdots\downarrow\downarrow\right\rangle =N(\frac{h^{x}}{-4J})\left\vert
\Psi_{i}\right\rangle .
\]

After the two fermions created, one of them moves along the Ising chain with
following steps,
\begin{align}
&  (\frac{h^{x}}{E_{0}-\hat{H}_{0}}\sum_{i}\sigma_{i}^{x})^{2}\left\vert
\downarrow\downarrow\cdots\downarrow\downarrow\right\rangle \nonumber\\
&  =(\frac{h^{x}}{E_{0}-\hat{H}_{0}}\sum_{i}\sigma_{i}^{x})N(\frac{h^{x}}
{4J})\left\vert \Psi_{i}\right\rangle \nonumber\\
&  =N(\frac{h^{x}}{4J})(\frac{h^{x}}{E_{0}-\hat{H}_{0}}\sum_{i}\sigma_{i}
^{x})\left\vert \Psi_{i}\right\rangle \nonumber\\
&  =N(\frac{h^{x}}{4J})(\frac{h^{x}}{E_{0}-\hat{H}_{0}}\sigma_{i+1}
^{x})\left\vert \Psi_{i}\right\rangle \nonumber\\
&  =N(\frac{h^{x}}{-4J})(\frac{h^{x}}{-4J})\left\vert \Psi_{i}^{\prime
}\right\rangle ,\nonumber
\end{align}
where $\left\vert \Psi_{i}^{\prime}\right\rangle $ is the excited state with
two fermions at sites $i+1$ and $i-1$. By this way, the fermion moves around
the chain step by step. When coming back to the starting point the fermion
annihilates with the partner fermion, which implies that the system changes
with the ground state flipped from $\left\vert \downarrow\downarrow
\cdots\downarrow\downarrow\right\rangle $ to $\mid\uparrow\uparrow
\cdots\uparrow\uparrow\rangle$. So we finally get the energy splitting
\begin{align}
\Delta E  &  =2\delta E=2U_{I}^{(N)}\nonumber\\
&  =2\langle\uparrow\uparrow\cdots\uparrow\uparrow\mid\hat{H_{I}}(\frac
{1}{E_{0}-\hat{H}_{0}}\hat{H_{I}})^{N-1}\left\vert \downarrow\downarrow
\cdots\downarrow\downarrow\right\rangle \nonumber\\
&  =2\times N\frac{(h^{x})^{N}}{(-4J)^{N-1}}=8NJ(\frac{h^{x}}{-4J})^{N}.
\end{align}

For the chain with open boundary condition, the tunneling process is much
different from our discussion above under the periodic boundary condition. In
this case, the dominant tunneling of a single virtual fermion is from one side
to the other with the tunneling splitting
\begin{equation}
\Delta E=2\delta E=2\frac{(h^{x})^{N}}{(2J)^{N-1}}. \label{delt}%
\end{equation}

By this high-order perturbative approach we have studied the tunneling
splitting between the two degenerate ground states. Fig. 3 shows the MQT
calculated numerically by the exact diagonalization technique, which indicates
that our theoretical results agree well with the numerical results. After
adding the perturbation field $h^{x}$ to the Ising chain, we are able to
generate two fermions and even drive one of them hopping around the chain by
considering high-order perturbation terms. In this way we could manipulate the
tunneling splitting of the two degenerate ground states. In addition, in the
sense of 'perturbative treatment', our approach can only be applied to the
cases under a small external field. Nevertheless, the tunneling induced by the
transverse disturbance field $h^{x}$ in our model makes it possible to realize
a string NOON state of the many-body state components $\left\vert
\uparrow\uparrow\cdots\uparrow\uparrow\right\rangle $ and $\left\vert
\downarrow\downarrow\cdots\downarrow\downarrow\right\rangle $.

\begin{figure}[ptb]
\includegraphics[width=0.5\textwidth]{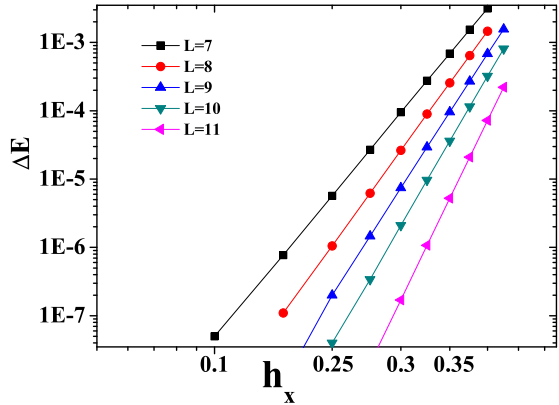}\caption{(Color online) The
energy splitting between the two degenerate ground states in an external field
along $x$-direction ($J=1$). Here $L$ denotes lattice the number. }%
\label{Fig.3}%
\end{figure}

Now we discuss how to generate a NOON state in a linear trapped-ion system. We
may employ $S_{1/2}$ $(m_{j}=-1/2)$ and $D_{5/2}$ $(m_{j}=-1/2)$
of$\ ^{40}\allowbreak Ca^{+}$ \cite{Ca} as the spin states $\left\vert
\uparrow\right\rangle $ and $\left\vert \downarrow\right\rangle $ in above
transverse Ising model, respectively, where the lifetime of $D_{5/2}$ of
$^{40}\allowbreak Ca^{+}$ ion is longer than $\allowbreak1\allowbreak$ $\sec$.
To carry out the present scheme in the ion trap, one should firstly confine
and cool $N$ $^{40}\allowbreak Ca^{+}$ ions in a linear Paul trap, and make
them located equidistantly in an array. Then we could generate a transverse
Ising model by coupling the internal states of the ions and their collective
motional phonon modes by choosing suitable values of the trap frequency and
Rabi frequency of the system \cite{ion1,onestep1,ion3}. Experimentally, such a
transverse Ising model has been simulated by trapped ions \cite{kimising}.
Furthermore, several other methods are reported in trapped-ion system to
create the $\left\vert NOON\right\rangle $ states, such as the methods in
\cite{ion1,onestep1} and more recently in \cite{ion3}.

In our case, after introducing the nearest-neighbor Ising interaction between
neighboring ions, the whole Ising chain should be polarized to one of the two
degenerate ferromagnetic ground states by irradiating the ion-chain with
specially polarized lasers. For example, we can initialize the Ising chain to
its ferromagnetic ground state $\left\vert \downarrow\downarrow\cdots
\downarrow\downarrow\right\rangle $ which corresponds to the $N$ spin-down of
the chain. The following task is to realize the transformation operator
$\hat{U_{I}}(0,-\infty)$ in the system. Note that the dynamical evolution
caused by the small transverse perturbation field $h_{x}$ can be expressed as
$N-1$-th order perturbation of $\hat{U_{I}}(0,-\infty)$, which is proportional
to the action of $e^{i\Delta W_{X}}$. As we had mentioned before, the purpose
we introduce $h_{x}$ is to induced the tunneling process between the two
degenerate ferromagnetic limit ground states $\left\vert \uparrow
\uparrow\cdots\uparrow\uparrow\right\rangle $ and $\left\vert \downarrow
\downarrow\cdots\downarrow\downarrow\right\rangle $. It implies that the
transverse field $h_{x}$ is just a little disturbance compared to the big
Ising coupling $J$ along $z$-direction. Based upon the consideration above, we
may globally flip the whole ions in the chain in an adiabatic way to generate
the perturbative string operator $W_{X}=\prod\limits_{i}\sigma_{i}^{x}$, which
can be complemented by applying a Gaussian beam with large width
experimentally. This key step takes time of $\tau=\frac{\hbar}{\Delta E}%
=\frac{(2J)^{N-1}}{2(h^{x})^{N}}\hbar$ \cite{time}, where $\Delta E$ can be
calculated using Eq.(\ref{deltE}) by high order degenerate perturbation
method. If we take a entanglement involving 14 ions as an example\cite{14},
the time we need for the Gaussian beam is on the order of $\tau\simeq10^{-12}$s.
The ferromagnetic domain wall in the chain implies that two
fermions are created and every fermion has a $2J$ excitation energy which is
much greater than $\Delta E$. In this way, we have achieved a NOON state
$\left\vert NOON\right\rangle =\alpha\left\vert \uparrow\uparrow\cdots
\uparrow\uparrow\right\rangle +\beta e^{i\phi}\left\vert \downarrow
\downarrow\cdots\downarrow\downarrow\right\rangle $ in the trapped ions under
this transverse Ising model, the three parameters: $\alpha$, $\beta$ and
$\phi$ could be determined by experimental measurements.

In summary, we have proposed a scheme to generate the NOON state of the two
degenerate ferromagnetic ground states in the transverse Ising model through
MQT process. This processes could be effectively manipulated by the
perturbative field which is experimentally feasible in ion-trap systems. We
think that this Ising string NOON state may have potential application in
quantum computation and even in topological quantum computation if we could
introduce topological protection to this Ising string.

\begin{acknowledgments}
This work is supported by NFSC Grant No. 11174035, No. 11004226, No. 11274351,
No.11147154, and the National Basic Research Program of China (973 Program)
under the grant No. 2011CB921803, 2012CB921704.
\end{acknowledgments}

\end{document}